# Synthesis and/or grafting of noble metal nanoparticles by microplasma and by atmospheric plasma torch

D. Merche[1], C. De Vos[1], J. Baneton[1], T. Dufour[2], S. Godet[3] and F. Reniers[1]

[1] *Service de Chimie Analytique et des Interfaces (CHANI), Université Libre de Bruxelles, Brussels, Belgium*
[2] *Laboratoire de Physique des Plasmas, Université Pierre & Marie Curie, Paris, France*
[3] *4MAT, Université Libre de Bruxelles, Brussels, Belgium*

**Abstract:** Plasmas at atmospheric pressure are presented as a simple, fast, and versatile tool for the synthesis or/and the grafting of noble metallic NPs (Au, Pt) on substrates. In this study, noble metal NPs are generated either by the reduction of a gold salt in an aqueous medium by microplasma or either by the decomposition of a platinum or gold organometallic in the post-discharge of an atmospheric plasma torch. The latter can also be used for the grafting of NPs from a commercial colloidal solution.

## 1. Introduction

Nanoparticles have attracted a great scientific attention as they are widely used due to their particular properties in new nanoscaled technologies including medicine, physic, optic and electronic domains [1]. Many techniques such as chemical synthesis by using reducing and capping agent [2], sonochemistry [1], ultraviolet-visible (UV-vis) irradiation [3], flame pyrolysis [4] and laser ablation [5] are developed to produce particles and control their size, shape and aggregation. More recently, the generation of plasma at atmospheric pressure has known a growing interest in a wide range of applications [6,7] such as organic or inorganic plasma deposition, etching, grafting, pollutant removal and, especially for the synthesis of NPs. We present plasma at atmospheric pressure as a simple, fast and versatile tool for the synthesis, or/and grafting of noble metal NPs on desired substrates. Microplasmas present specific advantages for the synthesis of nanomaterials due to their unique characteristics: stability and non-thermal operations under atmospheric pressure due to the confinement. More recently, they were coupled with liquids, enabling applications in water treatment, medicine and material synthesis [8, 9]. In this study, NPs were synthesized on the one hand by microplasma from the reduction of a gold salt in aqueous medium, and on the other hand by the decomposition of a gold or platinum organometallic in the post-discharge of a RF plasma torch. The torch enables also to graft NPs from a colloidal solution while maintaining the size of the NPs of the starting solution [10-13]. Au as Pt NPs exhibit some interests in sensor [14], actuation [15] and catalysis [16] domains. Moreover gold has a remarkable antimicrobial activity [1].

## 2. Experiments and results

### 2.1 Au NPs synthesis by microplasma

A micro-scale plasma [17], as depicted in Fig. 1, was generated between a DC hollow capillary (0.6 mm inner diameter) and the surface of an aqueous solution. A platinum foil immersed in the solution served as a counter-electrode (mass electrode) and the tip was positioned about 1 mm above the aqueous surface. The reduction of the metal cations from tetrachloroauric acid at the plasma-solution interface, in the presence of a stabilizing agent (polyvinyl alcohol- PVA, $1\%_m$) was immediately observed (strong color change due to the formation of Au NPs).

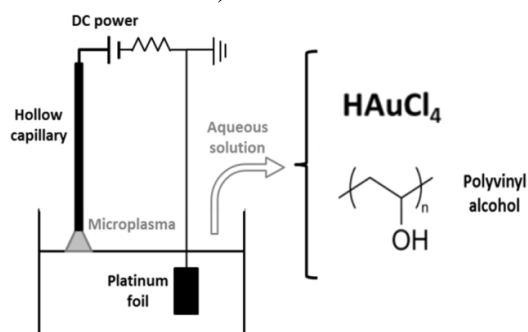

Fig. 1. Representation of the experimental device with the composition of the treated aqueous solution.

The as-synthesized NPs concentration in processed solutions was characterized by UV–vis absorbance spectroscopy, showing intense plasmon bands associated to spherical nanoparticles at 530 nm. According to the Mie theory [18], the bands of Surface Plasmon Resonance (SPR) bring information regarding the size, size distribution, shape, and agglomeration of the NPs. Those results were confirmed by Transmission

Electron Microscopy (TEM). TEM pictures (Fig.2) attest that decreasing the concentration of metallic salts lead to spherical and unaggregated particles characterized by a sharp distribution. For example, a concentration of 0.2 mM of $HAuCl_4$ leads to a large amount of Au particles of $8 \pm 3$ nm.

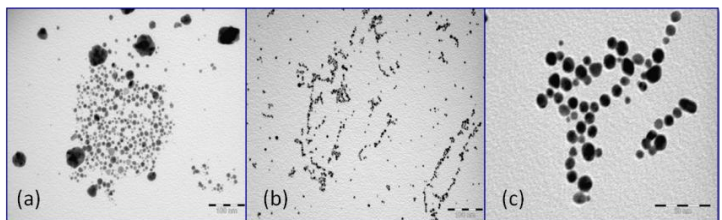

Fig. 2. TEM images of Au NPs synthesized from $HAuCl_4.3H_2O$ (a) 1.0 mM and (b,c) 0.2 mM solutions in presence of PVA stabilizing agent ($1\%_m$).

X-ray photoelectron spectroscopy (XPS) allows pointing out the metallic nature of the NPs (double peaks at 83.6 and 87.1 eV corresponding to Au $4f_{7/2}$ and Au $4f_{5/2}$, respectively). The dependence of the charge injected by the microplasma by varying the discharge current (from 2 to 10 mA) and process time (5 to 15 min) was studied monitoring the UV-vis absorbance of the solution after plasma treatment. Fig. 3 reveals a near linear dependence of the NPs concentration with the injected charge.

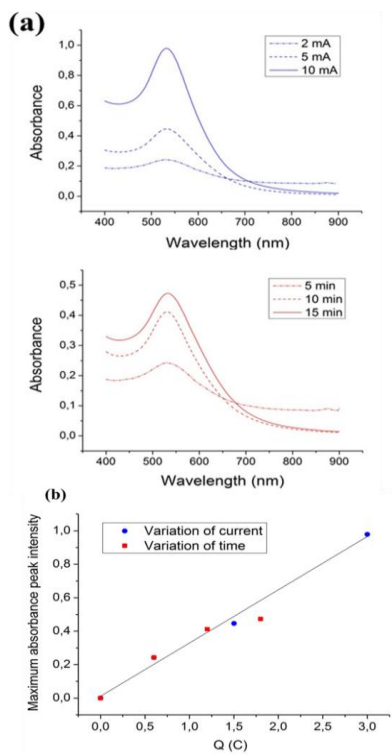

Fig. 3. (a) UV–vis absorbance spectra of solution of $HAuCl_4$ reduced by a microplasma electrode at different times (5–15 min) and discharge currents (2–10 mA) and (b) maximum absorbance peak intensity of the SPR band at ~530 nm versus the total charge injected, $Q$.

The colloidal solution generated by microplasma could be used for the subsequent grafting of the NPS on desired substrate using a plasma torch as described in 2.4.

### 2.2 Au NPs synthesis and grafting from organometallic by RF plasma torch

Au NPs were synthesized from the decomposition of gold acetate ($Au(O_2CCH_3)_3$) in the post-discharge of an RF plasma torch using Ar as the vector gas (30 L/min) in an open air environment. A gold acetate solution of 0.6 mM in acetone was nebulized out of the post-discharge on silicon and on carbon black substrates (pressed on a copper tape) after an activation phase to generate anchoring sites (oxidation, structural surface defects, etc.) for the NPs nucleation. The sample is exposed to the post-discharge after each pulverization(up to 30 pulverizations). The XPS surveys on treated Si wafers point out up to 14% of gold on the surface. As expected, the content of gold increases with the number of pulverizations of the solution nebulized on the substrate. The narrow spectra on the Au 4f region highlight that the gold is under metallic form. After ultrasonication in an acetone bath, 12% of (metallic) gold is still present on the Si surface, as shown in Fig.4, indicating that Au NPs are well grafted on the surface.

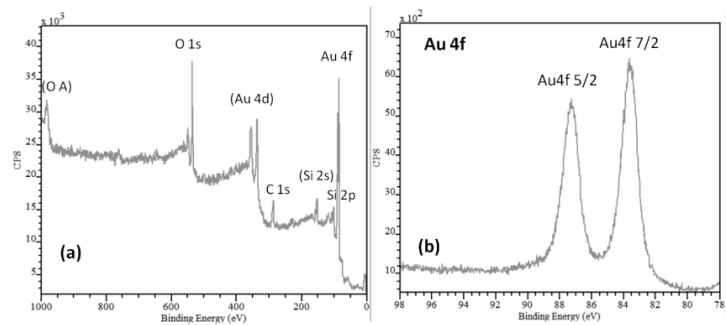

Fig. 4. (a) XPS (non-monochromatic Al anode) survey and (b) Au 4f spectra of gold NPs grafted on Si wafer (30 pulverizations), after ultrasonication in an acetone bath.

The morphology, the size and distribution size of Au NPs are observed by electron microscopy (SEM and TEM).

### 2.3 Pt NPs synthesis and grafting from organometallic by RF plasma torch

Pt NPs were generated from the decomposition of Platinum (II) acetylacetonate [$(C_5H_7O_2)^-_2 Pt^{2+}$], named ''[Pt(acac)$_2$], in the post-discharge of the previously described atmospheric plasma torch. The NPs were produced either by using nebulization (after an activation step) on silicon or carbon substrates, either by treating a powder mixture of carbon black and [Pt(acac)$_2$] by the post-discharge. The dried powder mixture (with a platinum weight ratio of 16 wt%) was pressed on a copper tape for a fundamental study [19].The optimal treatment time of the powder mixture to produce a significant amount of Pt NPs on the substrate, while minimizing the quantity of plasmagen gas, was evaluated by means of XPS. This study was also performed to investigate whether the exposure time could influence the oxidation state of the NPs. Table 1 reports the relative elemental composition after exposition of the samples to the active species emerging from the torch.

Table 1. Elemental composition after exposition of the samples function of the treatment time.

| Time (s) | % Pt | % C | % N | % O |
|---|---|---|---|---|
| 0 | 0.3 ± 0.1 | 93.4 ± 1.2 | 0 | 6.3 ± 1.3 |
| 30 | 6.1 ± 0.8 | 77.4 ± 1.1 | 2.3 ± 1.5 | 14.2 ± 1.0 |
| 60 | 6.9 ± 0.7 | 75.4 ± 1.2 | 2.7 ± 0.6 | 15.0 ± 1.3 |
| 150 | 8.3 ± 0.7 | 74.2 ± 1.5 | 2.8 ± 0.3 | 14.7 ± 0.9 |
| 300 | 9.9 ± 2.3 | 71.5 ± 3.1 | 0.9 ± 0.9 | 17.7 ± 2.4 |
| 600 | 13.3 ± 3.0 | 68.1 ± 1.4 | 1.5 ± 1.5 | 17.1 ± 1.8 |

The XPS composition of the native mixture is close to the native carbon black (93.4% C and 6.6% O). The native powder mixture presents an amount of Pt as low as 0.3% while it increases to 6.1% after 30 s and up to 13.3% after 600 s of plasma exposure. According to the fitting of the Pt XPS peaks (Fig. 5), no Pt under metallic form is detected for the untreated mixture of CB/[(Pt(acac)$_2$], while the most intense component of the Pt for the sample treated by plasma is the Pt(0). 70% of the Pt is under its metallic form after 30 seconds of plasma treatment, and 78 % after 600 seconds.

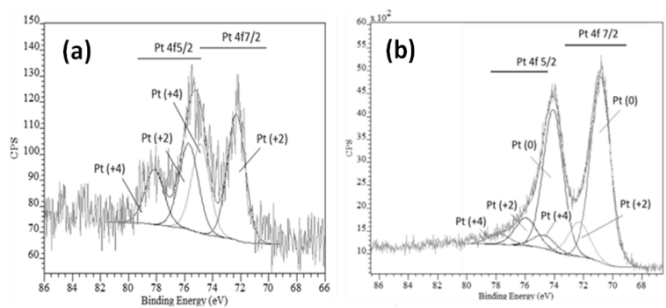

Fig. 5. Pt 4f XPS peaks for (a) CB/[Pt(acac)$_2$] untreated mixture (0 s), (b) mixture treated for 600 s.

In order to favor the degradation of the [Pt(acac)$_2$] organic component by a synergetic effect of the temperature during the plasma process, the influence of the temperature on the elemental composition was investigated. According to XPS results, for a surface temperature ranging from 333 K to 383 K, the Pt content increased from 9.9% to 18.0%, while it decreased to less than 0.7% for a temperature as high as 543 K (sublimation of Pt(acac)$_2$).

Electrodes for PEMFC applications showing correct electrochemical properties can be realized by treating the powder mixture overlaid (by spreading an ink) on gas diffusion layers (GDL) playing the role of water management [19]. SEM micrographs in Fig. 6 highlight the presence of NPs (average size close to 7 nm) grafted on carbon particles (forming micrometric aggregates presenting a bean shape with a typical diameter of 50 nm) overlaid on gas diffusion layers.

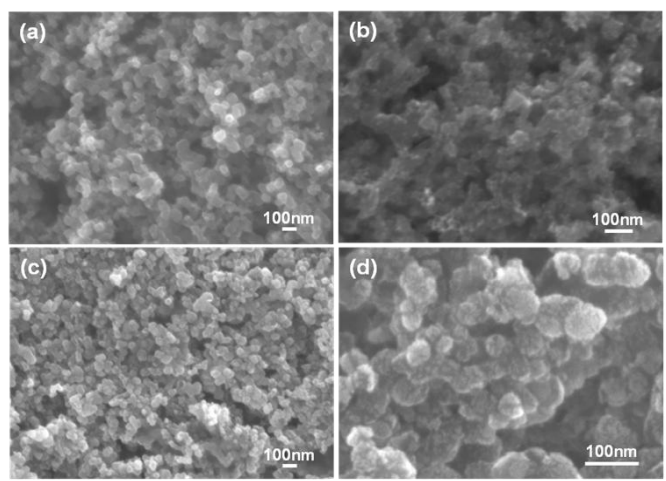

Fig. 6. SEM pictures of (a) the native Paxitech GDL (without catalyst), (b) the reference electrode (GDL with catalyst) and (c, d) the plasma electrode.

*2.4 Pt and Au NPs grafting from a colloidal solution by RF plasma torch*

Carbon substrates (carbon black and porous carbon Toray paper) and Si wafers were decorated by spraying a commercial platinum or a commercial gold colloidal solution after an activation step in the argon post-discharge of the atmospheric plasma torch [10-13].

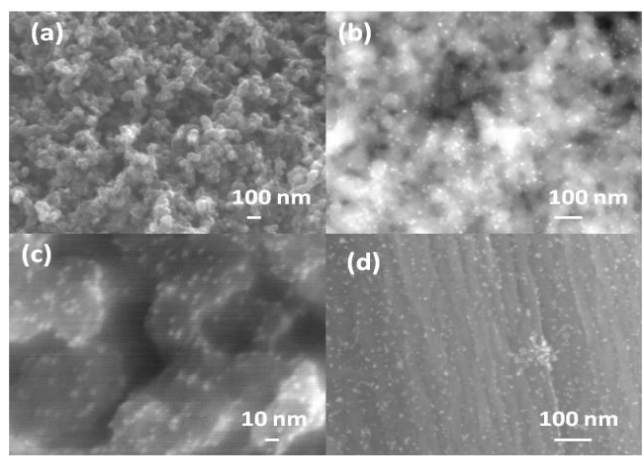

Fig. 7. SEM micrographs of Pt NPs grafted on (a,b,c) grains of CB (d) a fiber of a porous carbon Toray paper, by nebulization of a colloidal solution.

The NPs are still under metallic form and well grafted on the surface of the substrate as confirmed by the XPS results after ultrasonication in a methanol bath. SEM micrographs of Pt NPs grafted on carbon black and on carbon Toray Paper from the nebulization of a commercial colloidal solution have permit to observe the size and the size distribution of the NPs. As evidenced in Fig.7, NPs are well visible and homogeneously distributed on the surface. Moreover, they are sparsely or no agglomerated and their mean size is about 4 nm. The size of the NPs from the colloidal starting solution is preserving.

## 3. Conclusions

Plasmas at atmospheric pressure are of a tremendous interest since they are cheaper and easier to implement than plasma under vacuum. They are here introduced as a simple, multi-purpose, low-cost and rapid technique for the synthesis and/ or grafting of nanomaterials. Noble metal NPs were generated by means of different methods.
1) Microplasmas were used to reduce a gold salt at the surface of an aqueous solution. The gold salt concentration, the presence of a stabilizing agent (PVA), and the charge injected play a primordial role on the concentration, the diameter and the size distribution of the metallic NPs. Although the Au NPs can be generated without stabilizing agent, it permits to obtain smaller NPs and a sharp distribution size. By decreasing the salt to a concentration of 0.2 mM, NPs of about 8 nm can be created.
2) NPs were also synthesized by the decomposition of a platinum or gold organometallic in the post-discharge of an atmospheric plasma torch by nebulization of an organometallic solution or by treating a powder mixture CB/organometallic in one-pot. By these methods, the NPs are synthesized but also grafted on the support contrarily to the microplasma method. Pt NPs, mainly under metallic form, of about 7 nm were anchored on various supports (CB, Si wafers, etc.) and present a good quality of grafting as confirmed by XPS after ultrasonication of the hybrid materials. By varying the treatment time and the temperature for the powder mixture, the Pt content can be tuned. The grafting of Pt NPs on CB are of a huge interest for fuel cells applications.
3) The torch is also appropriated for the anchoring of NPs from a colloidal solution while maintaining the size of the NPs of the starting solution. This could be applied to the colloidal solutions obtained by microplasma.

## Acknowledgements

The authors would like to thanks the Walloon Region (Hylife and Cleanair projects) and the IAP Physical Chemistry of plasma surface interactions (Research project P7/34) funded by the Belgian Science Policy, for their financial support.